\input{aipcheck}

\documentclass[
    ,final            
  ]
  {aipproc}

\layoutstyle{8x11single}

\begin{document}

\title{\bf Cosmology-Independent Distance Moduli of 42 Gamma-Ray Bursts between Redshift of 1.44 and 6.60}

\classification{98.80.Es,  98.70.Rz}
 \keywords      { Observational cosmology  ---  gamma-ray bursts}

\author{Nan~Liang}{
  address={Department of Physics and Center for
Astrophysics, Tsinghua University, Beijing 100084, China} }

\author{ Shuang~Nan~Zhang}{
  address={Department of Physics and Center for
Astrophysics, Tsinghua University, Beijing 100084, China}
}

\begin{abstract}
This report is an update and extension of our paper accepted for
publication in ApJ (arXiv:0802.4262). Since objects at the same
redshift should have the same luminosity distance and the distance
moduli of type Ia supernovae (SNe Ia) obtained directly from
observations are completely cosmology independent, we obtain the
distance modulus of a gamma-ray burst (GRB) at a given redshift by
interpolating or iterating from the Hubble diagram of SNe Ia. Then
we calibrate five GRB relations without assuming a particular
cosmological model, from different regression methods, and construct
the GRB Hubble diagram to constrain cosmological parameters. Based
upon these relations we list the cosmology-independent distance
moduli of 42 GRBs between redshift of 1.44 and 6.60, with the
1-$\sigma$ uncertainties of 1-3\%.
\end{abstract}

\maketitle

\section{Introduction}\label{sec1}
Gamma-ray burst (GRB) luminosity/energy relations are connections
between measurable properties of the prompt gamma-ray emission with
the luminosity or energy. In recent years, several empirical GRB
luminosity relations as standard candles for cosmology research at
very high redshift have been proposed  \cite{GRB2008},
\cite{Schaefer2007}. An important concern in the application of GRBs
to cosmology is the dependence on the cosmological model in the
calibration of GRB relations in many works. For the difficulty to
calibrate the relations with a low-redshift sample, GRB relations
have usually been calibrated by assuming a particular cosmological
model. Therefore the circularity problem can not be avoided easily.
Many previous works treated the circularity problem by means of
statistical approaches. However, we note that the circularity
problem is not circumvented completely by means of statistical
approaches, because a particular cosmology model is required in
doing the joint fitting.

In this report, we present two new methods (interpolation method and
iterative method) to calibrate the GRB relations in a cosmological
model-independent way. This report is an update and extension of our
paper accepted for publication in ApJ\cite{Liang2008}. It is obvious
that objects at the same redshift should have the same luminosity
distance in any cosmology. There are so many SNe Ia that we can
obtain the luminosity distance (also the distance moduli) at any
redshift in the redshift range of SNe Ia by interpolating from SN Ia
data. Recently, on the basis of smoothing the noise of supernova
data over redshift, the authors in
\cite{Shafieloo2006},\cite{Wu2008} suggested a non-parametric method
in a model independent manner to reconstruct the luminosity distance
at any redshift in the redshift range of SNe Ia by the iterative
method. Furthermore, the luminosity distance of SNe Ia obtained
directly from observations are completely cosmological model
independent. Therefore, we can obtain the distance moduli of GRBs in
the redshift range of SNe Ia and calibrate GRB relations in a
completely cosmological model independent way and use the standard
Hubble diagram method to constrain the cosmological parameters from
GRB data at high redshift obtained by utilizing the relations.

\section{cosmological model independent High-z GRB distance moduli}
We first calibrate five GRB luminosity/energy
relations with the sample at $z\le1.4$, i.e., the luminosity
($L$)-spectral lag ($\tau_{\rm lag}$) relation \cite{Norris2000},
the $L$-variability ($V$) relation  \cite{Fenimore2000}, the
$L$-$E_{\rm p}$ relation \cite{Schaefer2003}, the
collimation-corrected energy ($E_{\gamma}$)-$E_{\rm p}$ relation
\cite{Ghirlanda2004}, the $\tau_{\rm RT}$-$L$ relation
\cite{Schaefer2007}, where $\tau_{\rm RT}$ is the minimum rise time
in the GRB light curve. A GRB luminosity relation can be generally
written in the form of
\begin{eqnarray}
\log y=a+b\log x,
\end{eqnarray}
where $a$ and $b$ are the intercept and slope of the relation
respectively; $y$ is the luminosity ($L$ in units of $\rm erg\
s^{-1}$) or the collimation-corrected energy ($E_{\gamma}$ in units
of $\rm erg$); $x$ is the GRB parameters measured in the rest frame,
e.g., $\tau_{\rm lag}(1+z)^{-1}/(0.1\rm \ s)$, $V(1+z)/0.02$,
$E_{\rm p}(1+z)/(300\ \rm keV)$, $\tau_{\rm RT}(1+z)^{-1}/\rm (0.1\
s)$. (We adopt the data of these quantities from Ref.
\cite{Schaefer2007}.)

We adopt the data of 192 SNe Ia \cite{Davis2007} and show them in
Figure 1. There is only one SN Ia point (the redshift of SN1997ff is
$z=1.755$) at $z>1.4$, therefore we exclude it from our SN Ia sample
used to interpolate the distance moduli of GRBs in the redshift
range of SN Ia sample.

\begin{figure}
 \begin{minipage}[b]{.6\textwidth}
     \centering

\includegraphics[height=.3\textheight]{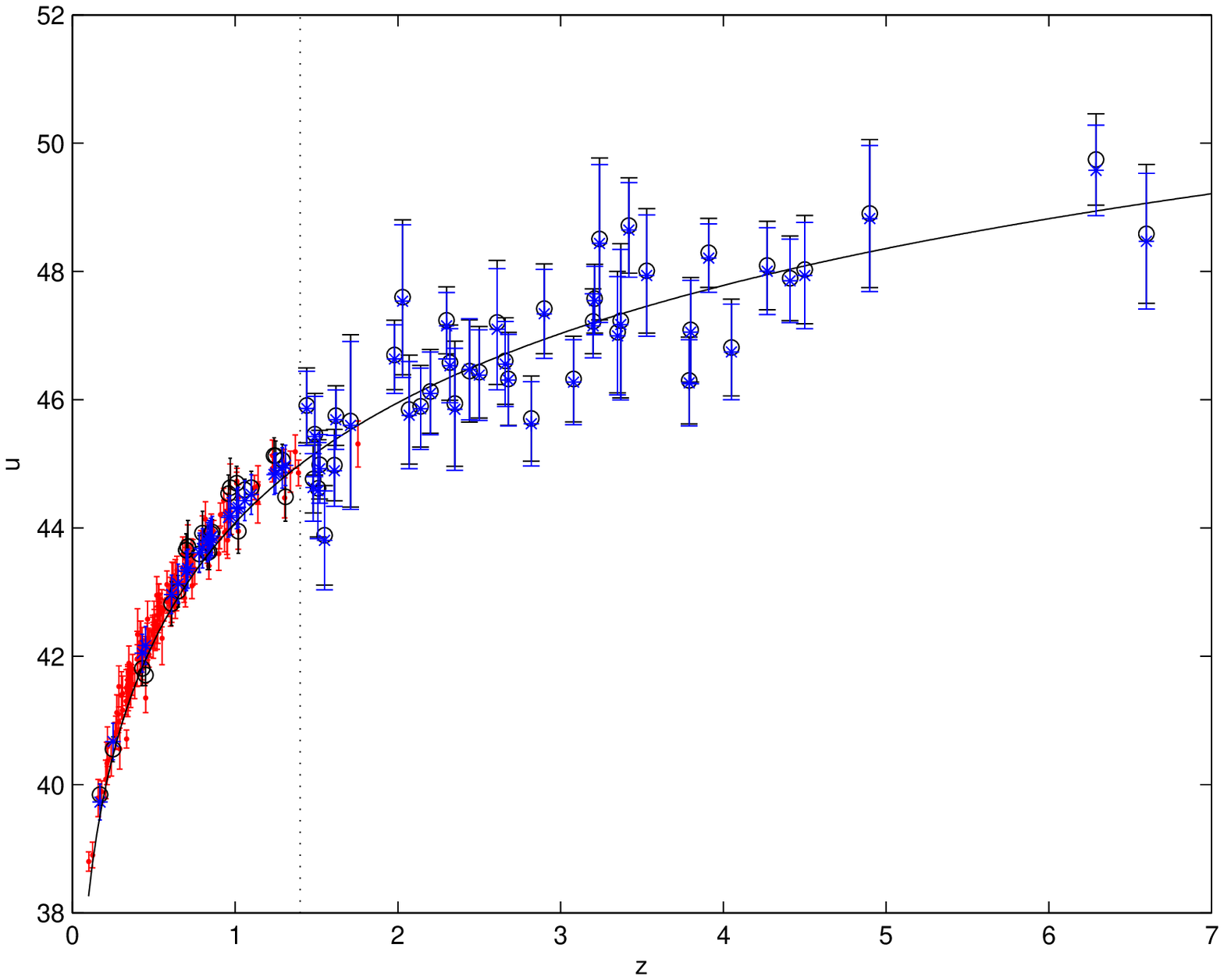}
   \end{minipage}%
   \begin{minipage}[b]{.4\textwidth}

 {\small {\bf FIGURE 1.} The Hubble Diagram of 192 SNe Ia (red dots) and the 69 GRBs
obtained by using the cosmology-independent  methods. The 27 GRBs at
$z\le1.4$ are obtained by using the interpolation and iterative
methods directly from SN Ia data; and the 42 GRBs at $z>1.4$ are
obtained by utilizing the five relations calibrated with the sample
at $z\le1.4$ using  by the cosmology-independent methods (black
circles: the interpolation method; blue stars: the iterative
method). The curve is the theoretical distance modulus in the
concordance model ($w_0= -1$, $\Omega_{\rm M}= 0.27$,
$\Omega_{\Lambda}= 0.73$), and the vertical dotted line represents
$z=1.4$.}
  \end{minipage}

 \end{figure}

Following a well known procedure in the analysis of large scale
structure, Shafieloo et al \cite{Shafieloo2006} used a Gaussian
smoothing function rather than the top hat smoothing function to
smooth the noise of the Sne Ia data directly. In order to obtain
important information on interesting cosmological parameters
expediently, when doing the Gaussian smoothing\cite{Shafieloo2006}
$\ln d_L(z)$, rather than the luminosity distance $d_L(z)$ or
distance modulus $\mu(z)$, is studied by the iterative method. We
thus follow the iterative procedure and adopt the results from Ref.
\cite{Wu2008},
\begin{eqnarray}
\ln d_L(z)^s_n=\ln d_L(z)^s_{n-1}+N(z)\sum_i(\ln d^{obs}_L(z_i)-\ln
d_L(z_i)^s_{n-1})\exp\big[-\big(\ln^2\big((1+z)/(1+z_i)\big)\big)/\big(
2\triangle^2\big)\big],
\end{eqnarray}
with a normalization parameter $N(z)^{-1}=\sum_i
\exp\big[-\big(\ln^2\big((1+z)/(1+z_i)\big)\big)/\big(
2\triangle^2\big)\big]$, and $\triangle=0.6$.  $d_L(z)^s_n$
represents the smoothed luminosity distance at any redshift $z$
after the $n$th iteration and $d_L(z)^s_{0}$ denotes a guess
background model. It has been shown that the results are not
sensitive to the chosen vaule of $\Delta$ and the assumed initial
guess model. $ d^{obs}_L (z_i)$ is the observed one from the SN Ia
data. The best fitting result is obtained by minimizing
$\chi^2_n=\sum_i
(\mu(z_i)_n-\mu^{obs}(z_i))^2/\sigma^2_{\mu_{obs,i}}\;$.

The isotropic luminosity of a burst is calculated  by $L = 4\pi
d_{\rm L}^2 P_{\rm bolo}$, where $d_{\rm L}$ is the luminosity
distance of the burst and $P_{\rm bolo}$ is the bolometric flux of
gamma-rays in the burst. The isotropic energy released from a burst
is given by $ E_{\rm iso} = 4\pi d_{\rm L}^2 S_{\rm bolo}
(1+z)^{-1}, $ where $S_{\rm bolo}$ is the bolometric fluence of
gamma-rays in the burst at redshift $z$. The total
collimation-corrected energy is then calculated by $ E_{\gamma} =
F_{\rm beam}E_{\rm iso}, $ where the beaming factor, $F_{\rm beam}$
is $(1-\cos\theta _{\rm jet})$ with the jet opening angle
($\theta_{\rm jet}$), which is related to the break time ($t_{\rm
b}$).

We determine the values of the intercept ($a$) and the slope ($b$)
calibrated with the GRB sample at $z\le1.4$ by using the
interpolation\cite{Liang2008} and iterative\cite{Wu2008} methods,
respectively. We first use the same regression method as used in
Ref.\cite{Schaefer2007}: the bisector of the two ordinary
least-squares(OLS) linear regressions, which has been
 discussed in Ref.\cite{Isobe1990}, OLS(Y|X) (OLS regression of the
dependent variable Y against the independent variable X) and its
inverse OLS(X|Y). In order to avoid specifying ``dependent'' and
``independent'' variables,  the OLS(Y|X) and the OLS(X|Y) lines
should be bisected. The OLS regressions does not take the errors
into account; but the use of weighted least-squares(WLS), taking
into account the measurable uncertainties, results in almost
identical best fits. Therefore, the WLS bisector regression can be
obtained by bisecting the WLS(Y|X) and the WLS(X|Y) lines.  The
calibration results from two regression methods (the OLS bisector
and the WLS bisector) are summarized in Table 1 and we also list the
results calibrated with the same sample ($z\le1.4$) by assuming the
$\Lambda$CDM model for comparison.
From Table 1, we find that results from the two regression methods
make no significant difference from each other. Therefore,  taking
into account the measurement uncertainties in the regression, indeed
will not change the fitting parameters significantly, when the
measurement uncertainties are smaller than the intrinsic error
\cite{Schaefer2007}. We also find that results obtained by assuming
the $\Lambda$CDM with the same sample differ only slightly from, but
still fully consistent with those calibrated by using our
interpolation method. The reason for this is easy to understand,
since the $\Lambda$CDM is fully compatible with SN Ia data.
Nevertheless, it should be noticed that the calibration results
obtained by using the interpolation or iterative method directly
from SN Ia data are completely cosmology independent.


\begin{table}
\begin{tabular}{|l|cc|cc|cc|}
\hline

  & \tablehead{2}{|c|}{b}{Interpolation method}
  & \tablehead{2}{|c|}{b}{Iterative method}
  & \tablehead{2}{|c|}{b}{$\Lambda$CDM model}    \\

\hline
                                           &$a$&$b$&$a$&$b$&$a$&$b$\\
\hline
$\tau_{\rm lag}$-$L$ relation             &52.22$\pm0.09$& -1.07$\pm0.13$&52.22$\pm0.09$& -1.10$\pm0.14$&52.15$\pm0.10$& -1.11$\pm0.14$ \\
                                          &52.22$\pm0.09$& -1.07$\pm0.14$&52.22$\pm0.09$& -1.10$\pm0.15$&52.14$\pm0.10$& -1.10$\pm0.14$ \\

\hline
$V$-$L$ relation                          &52.59$\pm0.13$&  2.05$\pm0.27$&52.56$\pm0.13$&  2.03$\pm0.30$&52.48$\pm0.13$&  2.04$\pm0.30$ \\
                                          &52.59$\pm0.13$&  2.08$\pm0.37$&52.57$\pm0.13$&  2.10$\pm0.42$&52.49$\pm0.13$&  2.11$\pm0.42$ \\
\hline
$L$-$E_{\rm p}$ relation                  &52.26$\pm0.09$&  1.69$\pm0.11$&52.23$\pm0.09$&  1.65$\pm0.12$&52.15$\pm0.09$&  1.66$\pm0.12$ \\
                                          &52.26$\pm0.09$&  1.69$\pm0.16$&52.23$\pm0.09$&  1.65$\pm0.16$&52.15$\pm0.09$&  1.65$\pm0.16$ \\
\hline
$E_{\gamma}$-$E_{\rm p}$ relation         &50.71$\pm0.07$&  1.79$\pm0.18$&50.68$\pm0.06$&  1.75$\pm0.18$&50.59$\pm0.06$&  1.75$\pm0.19$ \\
                                          &50.71$\pm0.06$&  1.68$\pm0.29$&50.68$\pm0.06$&  1.63$\pm0.30$&50.60$\pm0.06$&  1.64$\pm0.30$ \\
\hline
$\tau_{\rm RT}$-$L$ relation              &52.64$\pm0.11$& -1.32$\pm0.15$&52.60$\pm0.10$& -1.30$\pm0.15$&52.52$\pm0.11$& -1.30$\pm0.15$ \\
                                          &52.64$\pm0.11$& -1.33$\pm0.16$&52.60$\pm0.10$& -1.31$\pm0.16$&52.52$\pm0.11$& -1.31$\pm0.17$ \\
\hline
\end{tabular}
\caption{Calibration results ($a$: intercept,  $b$: slope) with
their 1-$\sigma$ uncertainties for the five GRB luminosity/energy
relations with the sample at $z\le1.4$, by using  the interpolation
 and iterative methods directly from SNe Ia data, and by
assuming a particular cosmological models (the $\Lambda$CDM model).
Only the data available from Table 4 in Ref.\cite{Schaefer2007} are
used to calibrate the five GRB relations. For each relation, the
results on the first and second lines are calibrated from the OLS
and WLS bisector regressions, respectively. } \label{tab:a}
\end{table}

By utilizing the calibrated relations at high redshift ($z>1.4$), we
are able to obtain the luminosity ($L$) or energy ($E_\gamma$) of
each burst at $z>1.4$. We use the same method used in
Ref.\cite{Schaefer2007} to obtain the best estimate $\mu$ for each
GRB which is the weighted average of all available distance moduli.
The derived distance modulus for each GRB is
\begin{eqnarray}
\mu = (\sum_i\mu_{\rm i} / \sigma_{\mu_{\rm i}}^2)/(\sum_i
\sigma_{\mu_{\rm i}}^{-2}),
\end{eqnarray}
with its uncertainty $ \sigma_{\mu} = (\sum_i
\sigma_{\mu_{\rm i}}^{-2})^{-1/2}$, where the summations run from 1
to 5 over the five relations used in Schaefer (2007) with available
data.

We plot the Hubble diagram of the 69 GRBs obtained by using the
interpolation and iterative methods in Figure 1. The 27 GRBs at
$z\le1.4$ are obtained by using the interpolation and iterative
methods directly from SNe data. The distance moduli of these 42 GRBs
at $z>1.4$ are obtained by utilizing the five relations calibrated
with the sample at $z\le1.4$ using the cosmology-independent
methods. The derived distance moduli for the 42 GRBs ($z>1.4$) with
different methods are listed in Table 2, together with the average
values between different methods, which should be the least biased
and most robust values to be used to study the expansion history of
the Universe up to $z=6.60$. It should be noted that the 1-$\sigma$
uncertainties listed in Table 2, ranging between 1\% to 3\%, include
both the measurement uncertainties and intrinsic scattering in these
luminosity/energy relations.

\begin{table}
\begin{tabular}{cccccccc}
\hline
  & \tablehead{1}{c}{b}{GRB}
  & \tablehead{1}{c}{b}{$z$}
  & \tablehead{1}{c}{b}{$\mu_1^a$}
  & \tablehead{1}{c}{b}{$\mu_1^b$}
  & \tablehead{1}{c}{b}{$\mu_2^a$}
  & \tablehead{1}{c}{b}{$\mu_2^b$}
  & \tablehead{1}{c}{b}{$\mu$} \\
\hline

&   050318  &   1.44    &   45.86    $\pm$      0.58    &   45.91    $\pm$      0.59    &   45.80    $\pm$      0.57    &   45.86    $\pm$      0.58    &   45.86    $\pm$      0.58    \\
&   010222  &   1.48    &   44.87    $\pm$      0.51    &   44.77    $\pm$      0.53    &   44.73    $\pm$      0.51    &   44.63    $\pm$      0.53    &   44.75    $\pm$      0.52    \\
&   060418  &   1.49    &   45.46    $\pm$      0.63    &   45.46    $\pm$      0.63    &   45.43    $\pm$      0.63    &   45.43    $\pm$      0.63    &   45.45    $\pm$      0.63    \\
&   060502  &   1.51    &   44.61    $\pm$      0.75    &   44.62    $\pm$      0.76    &   44.57    $\pm$      0.75    &   44.58    $\pm$      0.75    &   44.60    $\pm$      0.75    \\
&   030328  &   1.52    &   44.99    $\pm$      0.54    &   44.99    $\pm$      0.54    &   44.92    $\pm$      0.54    &   44.92    $\pm$      0.53    &   44.96    $\pm$      0.54    \\
&   051111  &   1.55    &   43.90    $\pm$      0.77    &   43.89    $\pm$      0.77    &   43.82    $\pm$      0.77    &   43.81    $\pm$      0.77    &   43.85    $\pm$      0.77    \\
&   990123  &   1.61    &   45.15    $\pm$      0.53    &   44.98    $\pm$      0.56    &   45.06    $\pm$      0.53    &   44.89    $\pm$      0.55    &   45.02    $\pm$      0.54    \\
&   990510  &   1.62    &   45.76    $\pm$      0.47    &   45.75    $\pm$      0.47    &   45.69    $\pm$      0.46    &   45.69    $\pm$      0.46    &   45.72    $\pm$      0.46    \\
&   050802  &   1.71    &   45.67    $\pm$      1.34    &   45.67    $\pm$      1.34    &   45.60    $\pm$      1.30    &   45.60    $\pm$      1.31    &   45.64    $\pm$      1.32    \\
&   030226  &   1.98    &   46.70    $\pm$      0.54    &   46.70    $\pm$      0.54    &   46.63    $\pm$      0.54    &   46.64    $\pm$      0.54    &   46.67    $\pm$      0.54    \\
&   060108  &   2.03    &   47.60    $\pm$      1.21    &   47.60    $\pm$      1.21    &   47.54    $\pm$      1.19    &   47.54    $\pm$      1.19    &   47.57    $\pm$      1.20    \\
&   000926  &   2.07    &   45.84    $\pm$      0.85    &   45.85    $\pm$      0.85    &   45.74    $\pm$      0.83    &   45.76    $\pm$      0.84    &   45.80    $\pm$      0.84    \\
&   011211  &   2.14    &   45.85    $\pm$      0.64    &   45.90    $\pm$      0.64    &   45.81    $\pm$      0.63    &   45.86    $\pm$      0.63    &   45.85    $\pm$      0.63    \\
&   050922  &   2.20    &   46.13    $\pm$      0.65    &   46.13    $\pm$      0.65    &   46.10    $\pm$      0.64    &   46.10    $\pm$      0.65    &   46.12    $\pm$      0.65    \\
&   060124  &   2.30    &   47.29    $\pm$      0.52    &   47.24    $\pm$      0.52    &   47.20    $\pm$      0.51    &   47.15    $\pm$      0.52    &   47.22    $\pm$      0.52    \\
&   021004  &   2.32    &   46.58    $\pm$      0.59    &   46.58    $\pm$      0.59    &   46.53    $\pm$      0.58    &   46.53    $\pm$      0.58    &   46.56    $\pm$      0.59    \\
&   051109  &   2.35    &   45.94    $\pm$      0.97    &   45.94    $\pm$      0.98    &   45.85    $\pm$      0.95    &   45.85    $\pm$      0.95    &   45.90    $\pm$      0.96    \\
&   050406  &   2.44    &   46.45    $\pm$      0.80    &   46.45    $\pm$      0.80    &   46.48    $\pm$      0.79    &   46.47    $\pm$      0.79    &   46.46    $\pm$      0.79    \\
&   030115  &   2.50    &   46.43    $\pm$      0.71    &   46.43    $\pm$      0.71    &   46.39    $\pm$      0.70    &   46.38    $\pm$      0.71    &   46.41    $\pm$      0.71    \\
&   050820  &   2.61    &   47.22    $\pm$      0.96    &   47.21    $\pm$      0.97    &   47.11    $\pm$      0.94    &   47.10    $\pm$      0.94    &   47.16    $\pm$      0.95    \\
&   030429  &   2.66    &   46.56    $\pm$      0.67    &   46.60    $\pm$      0.67    &   46.50    $\pm$      0.66    &   46.56    $\pm$      0.66    &   46.56    $\pm$      0.67    \\
&   060604  &   2.68    &   46.31    $\pm$      0.72    &   46.33    $\pm$      0.72    &   46.29    $\pm$      0.71    &   46.31    $\pm$      0.71    &   46.31    $\pm$      0.72    \\
&   050603  &   2.82    &   45.70    $\pm$      0.66    &   45.71    $\pm$      0.67    &   45.61    $\pm$      0.65    &   45.62    $\pm$      0.66    &   45.66    $\pm$      0.66    \\
&   050401  &   2.90    &   47.41    $\pm$      0.70    &   47.42    $\pm$      0.70    &   47.32    $\pm$      0.69    &   47.34    $\pm$      0.69    &   47.37    $\pm$      0.70    \\
&   060607  &   3.08    &   46.32    $\pm$      0.67    &   46.32    $\pm$      0.67    &   46.27    $\pm$      0.66    &   46.28    $\pm$      0.66    &   46.30    $\pm$      0.66    \\
&   020124  &   3.20    &   47.23    $\pm$      0.51    &   47.22    $\pm$      0.50    &   47.16    $\pm$      0.50    &   47.15    $\pm$      0.50    &   47.19    $\pm$      0.50    \\
&   060526  &   3.21    &   47.52    $\pm$      0.53    &   47.58    $\pm$      0.54    &   47.49    $\pm$      0.53    &   47.55    $\pm$      0.53    &   47.53    $\pm$      0.53    \\
&   050319  &   3.24    &   48.49    $\pm$      1.26    &   48.51    $\pm$      1.26    &   48.41    $\pm$      1.22    &   48.44    $\pm$      1.23    &   48.46    $\pm$      1.24    \\
&   050908  &   3.35    &   47.05    $\pm$      0.95    &   47.05    $\pm$      0.95    &   47.00    $\pm$      0.92    &   47.00    $\pm$      0.92    &   47.03    $\pm$      0.93    \\
&   030323  &   3.37    &   47.23    $\pm$      1.20    &   47.23    $\pm$      1.20    &   47.17    $\pm$      1.17    &   47.17    $\pm$      1.17    &   47.20    $\pm$      1.19    \\
&   971214  &   3.42    &   48.71    $\pm$      0.74    &   48.72    $\pm$      0.74    &   48.63    $\pm$      0.73    &   48.65    $\pm$      0.74    &   48.67    $\pm$      0.74    \\
&   060115  &   3.53    &   48.01    $\pm$      0.97    &   48.01    $\pm$      0.97    &   47.93    $\pm$      0.95    &   47.94    $\pm$      0.95    &   47.97    $\pm$      0.96    \\
&   050502  &   3.79    &   46.34    $\pm$      0.67    &   46.30    $\pm$      0.68    &   46.30    $\pm$      0.67    &   46.26    $\pm$      0.67    &   46.30    $\pm$      0.67    \\
&   060605  &   3.80    &   47.09    $\pm$      0.81    &   47.09    $\pm$      0.82    &   47.06    $\pm$      0.80    &   47.05    $\pm$      0.81    &   47.07    $\pm$      0.81    \\
&   060210  &   3.91    &   48.35    $\pm$      0.53    &   48.29    $\pm$      0.54    &   48.27    $\pm$      0.53    &   48.21    $\pm$      0.53    &   48.28    $\pm$      0.53    \\
&   060206  &   4.05    &   46.82    $\pm$      0.75    &   46.81    $\pm$      0.76    &   46.75    $\pm$      0.74    &   46.75    $\pm$      0.75    &   46.78    $\pm$      0.75    \\
&   050505  &   4.27    &   48.08    $\pm$      0.69    &   48.10    $\pm$      0.69    &   47.99    $\pm$      0.68    &   48.00    $\pm$      0.68    &   48.04    $\pm$      0.69    \\
&   060223  &   4.41    &   47.89    $\pm$      0.66    &   47.89    $\pm$      0.66    &   47.85    $\pm$      0.65    &   47.85    $\pm$      0.65    &   47.87    $\pm$      0.66    \\
&   000131  &   4.50    &   48.02    $\pm$      0.84    &   48.03    $\pm$      0.84    &   47.92    $\pm$      0.83    &   47.94    $\pm$      0.83    &   47.98    $\pm$      0.84    \\
&   060510  &   4.90    &   48.90    $\pm$      1.15    &   48.90    $\pm$      1.16    &   48.82    $\pm$      1.13    &   48.83    $\pm$      1.14    &   48.86    $\pm$      1.14    \\
&   050904  &   6.29    &   49.91    $\pm$      0.65    &   49.75    $\pm$      0.71    &   49.74    $\pm$      0.65    &   49.58    $\pm$      0.71    &   49.74    $\pm$      0.68    \\
&   060116  &   6.60    &   48.59    $\pm$      1.08    &   48.59    $\pm$      1.08    &   48.48    $\pm$      1.05    &   48.47    $\pm$      1.06    &   48.53    $\pm$      1.07    \\

\hline
\end{tabular}
\caption{Cosmology-independent distance moduli of 42 GRB ($1.44 \le
z\le 6.60$). $\mu_1$ and $\mu_2$ are obtained by the interpolation
and iterative methods, respectively; $\mu^a$ and $\mu^b$ are
obtained from the OLS and WLS bisector regressions, respectively.
$\mu=(\mu_1^a+\mu_1^b+\mu_2^a+\mu_2^b)/4$ and
$\sigma_{\mu}=(\sigma_{\mu_1^a}+\sigma_{\mu_1^b}+\sigma_{\mu_2^a}+\sigma_{\mu_2^b})/4$,
which should be least biased and most robust for applications. The
1-$\sigma$ uncertainties include both the measurement uncertainties
and intrinsic scattering in these luminosity/energy relations.}
\label{tab:b}
\end{table}

In Figure 2 we show examples of cosmological parameter fitting by
the minimum $\chi^2$ method. Figure 2\emph{a} and \emph{b} show the
joint confidence regions for ($\Omega_{\rm M},\Omega_{\Lambda}$) in
the $\Lambda$CDM model from the distance moduli of these 42 GRBs
($z>1.4$) obtained by utilizing the five relations calibrated with
the sample at $z\le1.4$, using the interpolation and iterative
methods, respectively. Figure 2\emph{c} and \emph{d} represent the
contours of likelihood in the ($\Omega_{\rm M}, w_0$) plane in the
dark energy model with a constant $w_0$ for a flat universe. Here we
adopt $H_0=70\ \rm km\ s^{-1}Mpc^{-1}$. All fitted parameters are
listed in Table 3. We find that the fitting results from the OLS
 regression differ only slightly from, but still fully
consistent with those from the WLS regression.  But the use of WLS,
taking into account the measurement uncertainties in the regression,
results in almost identical best fits.

\addtocounter{figure}{1}
\begin{figure}
\includegraphics[height=.3\textheight]{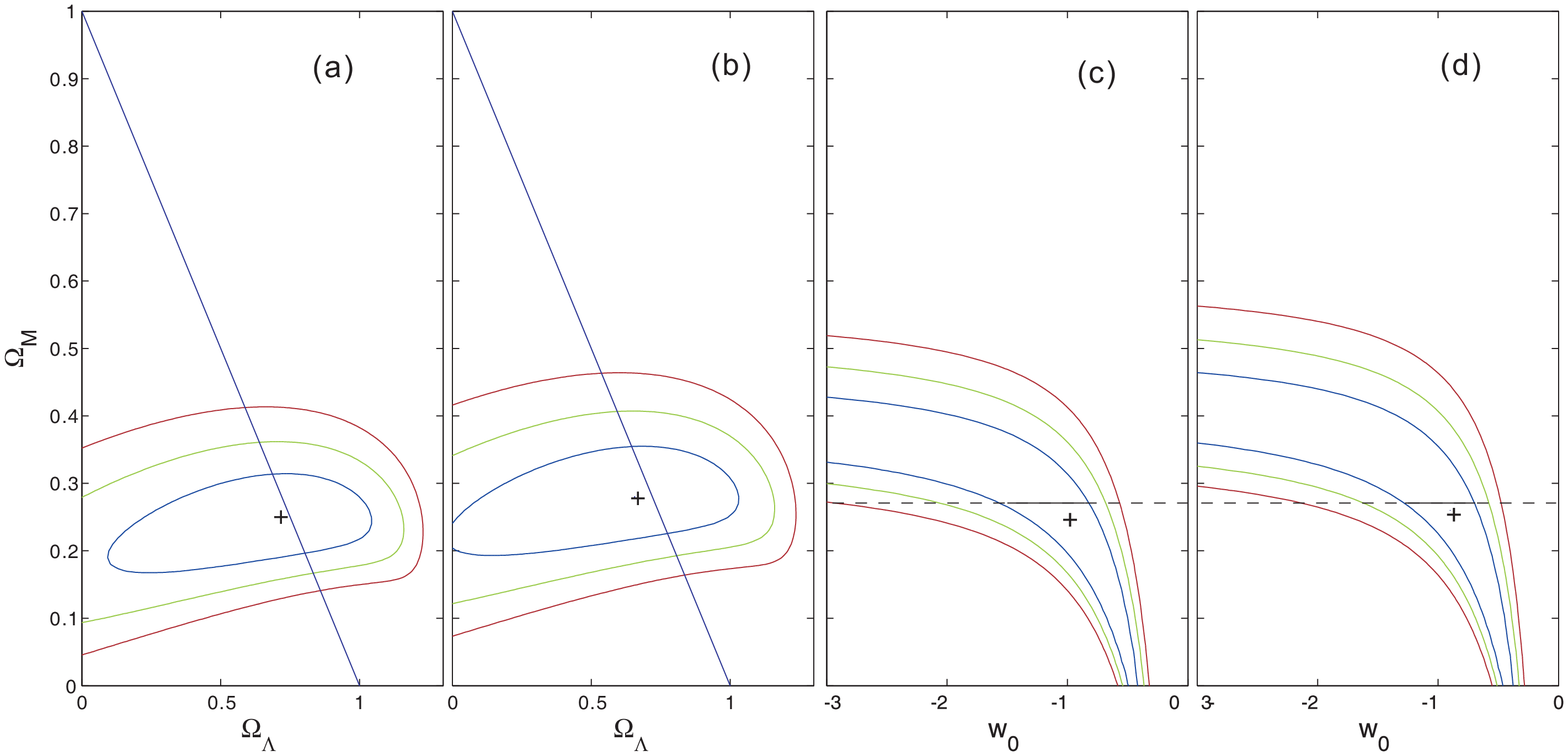}
\caption{(\emph{a}) and (\emph{b}) Joint confidence regions for
($\Omega_{\rm M}, \Omega_{\Lambda}$) in the $\Lambda$CDM model from
the data for 42 GRBs ($z>1.4$) obtained by utilizing the five
relations calibrated with the sample at $z\le1.4$, using the
interpolation and iterative methods, respectively. The dashed line
represents the flat universe. (\emph{c}) and (\emph{d}) Contours of
likelihood in the ($\Omega_{\rm M}, w_0$) plane in the dark energy
model with a constant $w_0$ for a flat universe. For each panel, the
horizontal line is for a prior of $\Omega_{\rm M}=0.27$. For all
four panels, the plus sign indicates the best-fit values and the
contours correspond to 1, 2, and 3-$\sigma$ confidence regions. The
numerical values of fitted parameters are listed in Table 3.}
\end{figure}

\begin{table}
\begin{tabular}{lcccccc}
\hline

  & \tablehead{1}{r}{b}{OLS(Y|X)}
  & \tablehead{1}{r}{b}{WLS(Y|X)}
   & \tablehead{1}{r}{b}{OLS(X|Y)}
  & \tablehead{1}{r}{b}{WLS(X|Y)}
   & \tablehead{1}{r}{b}{OLS Bisector}
    & \tablehead{1}{r}{b}{WLS Bisector}
   \\

\hline

$\Omega_{\rm M}$
&$0.35_{-0.05}^{+0.06}$&$0.35_{-0.06}^{+0.06}$ &$0.18_{-0.07}^{+0.05}$ &$0.21_{-0.07}^{+0.06}$&$0.27_{-0.06}^{+0.05}$&$0.28_{-0.06}^{+0.05}$\\
&$0.38_{-0.07}^{+0.07}$&$0.38_{-0.05}^{+0.07}$ &$0.21_{-0.08}^{+0.05}$ &$0.24_{-0.08}^{+0.06}$&$0.30_{-0.06}^{+0.06}$&$0.31_{-0.06}^{+0.06}$\\
\hline

$w_0$
&$-0.72_{-0.30}^{+0.21}$&$-0.73_{-0.31}^{+0.23}$&$-1.51_{-1.20}^{+0.50}$&$-1.29_{-0.96}^{+0.43}$&$-0.97_{-0.48}^{+0.28}$&$-0.94_{-0.45}^{+0.27}$\\
&$-0.63_{-0.25}^{+0.21}$&$-0.63_{-0.26}^{+0.20}$&$-1.25_{-0.75}^{+0.39}$&$-1.09_{-0.66}^{+0.35}$&$-0.84_{-0.36}^{+0.24}$&$-0.82_{-0.36}^{+0.24}$\\

\hline
\end{tabular}
\caption{The cosmological fitting results from the distance moduli
of these 42 GRBs ($z>1.4$) obtained utilizing the five relations
calibrated with the sample at $z\le1.4$ using the interpolation and
iterative methods, from different linear regressions. The fitted
$\Omega_{\rm M}$ is for the flat $\Lambda$CDM model. For the dark
energy model with a constant equation of state $w_0$
 is for a prior of $\Omega_{\rm
M}=0.27$.  For the every two lines of each parameter, the results on
the first line are derived from the interpolation method, and that
on the second line are from the iterative method.} \label{tab:c}
\end{table}

\section{Summary}
Since the distance modulus of any SN Ia is completely
cosmological model independent, we can obtain the distance modulus
of a GRB at a given redshift by interpolating or iterating from the
Hubble diagram of SNe Ia at $z\le1.4$, in order to calibrate the GRB
luminosity relations in a completely cosmology independent way.
Since our method does not depend on a particular cosmological model
when we calibrate the parameters of GRB luminosity relations, the
so-called circularity problem can be completely avoided. With this
method, we obtained the cosmology-independent distance moduli of 42
GRBs between redshift of 1.44 and 6.60, which are listed in Table 2.

With these cosmology-independent distance moduli of high redshift
GRBs, we construct the GRB Hubble diagram and constrain cosmological
parameters by the minimum $\chi^2$ method as in SN Ia cosmology. We
obtain $\Omega_{\rm M}=0.28_{-0.06}^{+0.05}$,
$\Omega_{\Lambda}=0.72_{-0.05}^{+0.06}$ for the flat $\Lambda$CDM
model from the GRB data obtained by using the interpolation method,
and $\Omega_{\rm M}=0.31_{-0.06}^{+0.06}$,
$\Omega_{\Lambda}=0.69_{-0.06}^{+0.06}$ from the data obtained by
using the iterative method. For the dark energy model with a
constant equation of state, we obtain $w_0=-0.94_{-0.45}^{+0.27}$
and $w_0=-0.82_{-0.36}^{+0.24}$ for a flat universe from the data
obtained by the two methods respectively, which is consistent with
the concordance model within the statistical error. Our result
suggests the the concordance model ($w_0=-1$, $\Omega_{\rm M}=0.27$,
$\Omega_{\Lambda}=0.73$) is still consistent with the GRB data at
higher redshift up to $z=6.6$.

\vspace{-3mm}
\begin{theacknowledgments}
\vspace{-2mm}
 We thank Hao Wei, Pu-Xun Wu, Yuan~Liu, Zi-Gao~Dai and En-Wei Liang for
kind help and discussions. This project was in part supported by the
Ministry of Education of China, Directional Research Project of the
Chinese Academy of Sciences under project KJCX2-YW-T03, by the
National Natural Science Foundation of China under grants 10521001,
10733010, and 10725313, and by 973 Program of China under grant
2009CB824800.
\end{theacknowledgments}



\bibliographystyle{aipproc}   


\end{document}